\documentstyle[twocolumn,prd,aps,epsf,floats]{revtex}
\begin{document}
\draft
\title{On Bubble Growth and Droplet Decay in Cosmological Phase Transitions}
\author{H.~Kurki-Suonio$^{a,}$\thanks{Email: hkurkisu@pcu.helsinki.fi}
and M.~Laine$^{b,c,}$\thanks{Email: m.laine@thphys.uni-heidelberg.de}}
\address{$^a$Research Institute for Theoretical Physics and
$^b$Department of Physics, \\
P.O.~Box 9, FIN-00014 University of Helsinki, Finland \\
$^c$Institut f\"ur Theoretische Physik,
Philosophenweg 16,
D-69120 Heidelberg, Germany}
\date{December 1, 1995}
\maketitle

\begin{abstract}
We study spherically symmetric bubble growth and droplet decay in
first order cosmological phase transitions, using a numerical code
including both the complete hydrodynamics of the problem and a
phenomenological model for the microscopic entropy producing mechanism
at the phase transition surface.  The small-scale effects of finite wall
width and surface tension are thus consistently incorporated. We verify
the existence of the different hydrodynamical growth modes proposed recently
and investigate the problem of a decaying quark droplet in the QCD
phase transition. We find that the decaying droplet leaves behind
no rarefaction wave, so that any baryon number inhomogeneity generated
previously should survive the decay.

\end{abstract}
\narrowtext

\vspace*{-11.3cm}
\noindent
\hspace*{9.5cm} \mbox{HU-TFT-95-71, HD-THEP-95-52, hep-ph/9512202}
\vspace*{9.8cm}

\section{Introduction}

The cosmological quark--hadron (QCD)
phase transition~\cite{w} took place when the age of the
universe was $t \sim 10^{-5}{\rm s}$.  The order of this transition remains
undetermined.  If it is of first order,
it may take up appreciable time. Unless
the latent heat is very small compared with the surface tension, the
probable sequence of events is:\\
1. Supercooling of the quark-gluon plasma.\\
2. Nucleation of bubbles of the hadron gas.\\
3. Rapid growth of hadron bubbles by deflagration.\\
4. Coupling in of particles without strong interactions.\\
5. Collisions of bubbles with shocks from other bubbles,
and reheating of the plasma to the critical temperature.\\
6. Slow growth of the hadron bubbles paced by the expansion of the universe,
leading to percolation.\\
7. Slow shrinking of remaining regions of quark plasma and their
becoming spherical.\\
8. Decoupling of particles without strong interactions.\\
9. Evaporation of the quark droplets.

The electroweak (EW) phase transition~\cite{kili} took place earlier,
at $t \sim 10^{-11}{\rm s}$. It is likely to be of first order,
since there is baryon asymmetry in the universe~\cite{krs}. However,
within the Standard Model, the order remains unclear for
Higgs masses $m_H\gtrsim 80$ GeV. If the transition is of first order,
it will likewise proceed through growing bubbles and shrinking
droplets, although the details will be different. For instance,
there may not be a stage of slow growth (6-7).

Clearly, to understand the possible physical consequences
of the cosmological phase transitions, all the stages should
be thoroughly understood. The nucleation period (1-2) from
the critical temperature $T_c$ to the nucleation temperature
$T_n$ seems to be rather well under control, and we do not touch it here.
For the QCD phase transition, the present understanding of the other
stages is the following: The general hydrodynamics of
bubble growth has been studied extensively~\cite{St,GKKM,%
KS,KajKur,MilPan,EIKR,Link,BonPan,HKLLM,KKT,La,H,KL}. Hydrodynamics
alone cannot determine the bubble solution, but due to
the strong coupling constant, there is not very much more
one can say~\cite{kk,IKKL}. The coupling in of
electromagnetic radiation
with the strongly interacting matter
at about the scale $10^4$ fm is discussed in~\cite{rm1,mr1},
the collisions with shocks in~\cite{KajKur,IKKL}, and the slow stages
at $T_c$ in~\cite{w}. In our opinion, the possible
formation of a shrinking similarity solution~\cite{RMP} at stage (7) is not
completely understood. Assuming that this happens, the decoupling
of electromagnetic radiation at the scale $10^4$ fm,
and the final stages until the quark droplet has a radius
of about 1 fm, have been discussed in~\cite{RMP,rm2}.
One should note that depending on the values of the latent
heat and surface tension, the phase transition might also
proceed rather differently from the standard scenario
assumed here~\cite{IKKL2}.

In this paper, we complete the hydrodynamic study of growing
bubbles (3) presented in~\cite{KL}, and investigate in detail
the final stages (9) from a radius of about 200 fm
until the time that the quark droplet has completely
disappeared, thus improving on~\cite{RMP}.
The final stages are important since
they determine the baryon number inhomogeneity
resulting from the QCD phase transition.

For the EW phase transition, there are a number
of microscopic investigations of bubble
growth~\cite{t,DHLLL,LMT,Kh,a,mp1}. These assume
that the change of temperature across the phase transition
surface, $\delta T/(T_c-T_n)$, is vanishing, and hence are
applicable only for very weak deflagrations. A phenomenological
investigation with one free parameter but without any
limitations on $\delta T/(T_c-T_n)$ is in~\cite{IKKL}.
Recently there has appeared also a microscopic investigation
with hydrodynamics included~\cite{mp2}, so that stage (3)
seems to be reasonably well under control.
As to the later stages, in~\cite{H}
it was argued that the collisions with the shock fronts (5)
might have interesting effects, if the bubble walls are slowed
down appreciably. This could happen if the temperature rises
close to $T_c$. Based on simple perturbative estimates, one often
assumes that reheating to $T_c$ and the stages (6-7) do not
exist; however, this need not necessarily be
the case, see e.g.~\cite{desy}.
In particular, lattice simulations~\cite{klrs} suggest
that the surface tension may be smaller than the perturbative
value, leading to smaller supercooling and possibly to
a reheating to~$T_c$. Then the final stages would proceed
similarly to the QCD case. The results of this paper
are applicable to the general characteristics of
bubble growth (3) and to the final stages (9).

We study the growth of bubbles and the decay of droplets in
first-order phase transitions as a hydrodynamical problem.  We
shall use the quark/hadron terminology, but as stated, our results
are qualitative and apply to some extent also to the electroweak
phase transition. A major uncertainty in these studies is the phase
boundary, the bubble or droplet wall, which remains poorly understood,
especially for QCD. Hence one usually has to resort to some kind of
parametrization of its properties.  We use
a cosmic-fluid--order-parameter-field model, which
allows simultaneous treatment of the phase boundary and surrounding
hydrodynamics within a single framework~\cite{IKKL}.  We have written a
hydrodynamical code using this model;  an earlier version was
plane-symmetric, but we now have a spherically symmetric version.
In the next Section, we describe the model and how
it differs from the one in~\cite{RMP,rm2}.
In Section~\ref{bubble} we investigate bubble growth
with our model, and in Section~\ref{droplet} droplet decay.

\section{The model}
\label{model}

We summarize here our cosmic-fluid--order-parameter-field model~\cite{IKKL}.
We ignore the chemical potential related to the small baryon number.
The local state of matter is then described by three
quantities: a local temperature~$T$,
a local flow velocity $u^\mu$ and a local value for the order
parameter $\phi$.  The order parameter has a temperature-dependent
effective potential $V(\phi,T)$.  The (meta)stable states of the
system are defined by the minima of the effective potential.

The equation of state is
\begin{eqnarray}
\epsilon(\phi,T) & = & 3aT^4 + V(\phi,T) - T\frac{\partial V}{\partial T}, \\
p(\phi,T)        & = &  aT^4 - V(\phi,T).
\end{eqnarray}
where $a=(\pi^2/90)g_*$.
The energy-momentum tensor is
\begin{equation}
T^{\mu\nu} = \partial^\mu\phi\partial^\nu\phi - g^{\mu\nu} \bigl(\frac{1}{2}
\partial_\alpha\phi\partial^\alpha\phi) + wu^\mu u^\nu + g^{\mu\nu}p,
\end{equation}
where $w \equiv \epsilon + p$ and
the metric convention is (--+++).
The energy-momentum conservation equation
$T^{\mu\nu}_{;\mu} = 0$ is split into two parts:
\begin{eqnarray}
T^{\mu\nu}_{;\mu}({\rm field}) & = & \phi^{;\mu}_{;\mu}\phi^{,\nu}
   - \frac{\partial V}{\partial\phi}\phi^{,\nu}
   = \eta u^{\mu}\phi_{,\mu}\phi^{,\nu}, \label{eqforfield}\\
T^{\mu\nu}_{;\mu}({\rm fluid}) & = & (wu^\mu u^\nu)_{;\mu} + p^{,\nu}
   + \frac{\partial V}{\partial\phi}\phi^{,\nu}
   = - \eta u^{\mu}\phi_{,\mu}\phi^{,\nu},
\label{eqforZ}
\end{eqnarray}
where ``;'' is a covariant derivative.
Here $\eta$ is a dissipative constant
relating entropy production to the gradients of~$\phi$
through $T(su^\mu)_{;\mu}=\eta(u^\mu\phi_{,\mu})^2$.
Eq.~(\ref{eqforfield}) contains the equation for the $\phi$-field,
\begin{equation}
\phi^{;\mu}_{;\mu} - \frac{\partial V}{\partial\phi} = \eta u^{\mu}\phi_{,\mu}.
\label{eom}
\end{equation}
Eq.~(\ref{eom}) (or its equivalent)
could in principle be derived from field theory
at least for the EW case,
see e.g.~\cite{mp2}, but we do not here attempt to do so.

In the spherically symmetric case, eq.~(\ref{eqforZ}) contains two
independent equations. For numerical hydrodynamics it is best to choose
eq.~(\ref{eqforZ}) with $\nu=r$, together with the contraction of
eq.~(\ref{eqforZ}) with $u_\nu$:
\begin{equation}
(\epsilon u^\mu)_{;\mu} + p u^\mu_{;\mu} -
\frac{\partial V}{\partial\phi} u^\mu
\phi_{,\mu} =  \eta(u^\mu\phi_{,\mu})^2 \label{eqforE}.
\end{equation}
{}From eqs.~(\ref{eqforZ}), (\ref{eom}), (\ref{eqforE}),
the final spherically symmetric equations in Minkowski space
then become
\begin{eqnarray}
-\partial_t^2\phi &+& \frac{1}{r^2}\partial_r(r^2\partial_r\phi) -
\frac{\partial V}{\partial\phi}
  =  \eta\gamma(\partial_t\phi + v\partial_r\phi), \\
\partial_t E &+& \frac{1}{r^2}\partial_r(r^2 Ev)
  + p\bigl[\partial_t\gamma + \frac{1}{r^2}\partial_r(r^2\gamma v)\bigr]
\nonumber \\
 & &  - \frac{\partial V}{\partial\phi}\gamma
(\partial_t\phi + v\partial_r\phi) =
\eta\gamma^2(\partial_t\phi + v\partial_r\phi)^2, \\
\partial_t Z &+& \frac{1}{r^2}\partial_r(r^2 Zv) +
\partial_r p + \frac{\partial V}{\partial\phi}
\partial_r\phi \nonumber \\
 & & =  -\eta\gamma(\partial_t\phi + v\partial_r\phi)\partial_r\phi,
\end{eqnarray}
where $E \equiv \epsilon\gamma$ and $Z \equiv w\gamma^2v$.

For the effective potential
$V(\phi,T)$ we use the simple parametrization
\begin{equation}
V(\phi,T) = \frac{1}{2}\gamma(T^2-T_0^2)\phi^2 - \frac{1}{3}\alpha T\phi^3 +
   \frac{1}{4}\lambda\phi^4
\label{VphiT}
\end{equation}
where~\cite{EIKR,kk}
\begin{eqnarray}
T_0 & = & \frac{T_c}{\sqrt{1+6\sigma/(L l_c)}}, \\
\alpha & = & \frac{1}{\sqrt{2\sigma l_c^5 T_c^2/3}}, \\
\gamma & = & \frac{L + 6\sigma / l_c}{6\sigma l_c T_c^2}, \\
\lambda & = & \frac{1}{3\sigma l_c^3}.
\end{eqnarray}
Here $\sigma$ is the surface tension, $l_c$ is the correlation length,
$L$ is the latent heat, and $T_c$ is the critical temperature of the
transition. We use the functional form in eq.~(\ref{VphiT}) due
to its simplicity; our results are qualitative and we expect
them to remain the same for any $V(\phi,T)$ including a first
order phase transition.

Let us now compare our model with that used
in~\cite{RMP,rm2} (RMP,RM) for studying droplet decay.
The main difference is that in our model the phase transition
surface has a microscopic structure and a finite width, whereas
in~\cite{RMP,rm2} it is a discontinuity, across
which explicit jump conditions are imposed.
For this reason RMP have to stop the evolution of the
decaying droplet at a radius of about
1 fm when the width of the interface starts to
have significance, whereas we are able to follow the decay until
the very end and beyond. We also believe that in our code the velocity
determination by the coefficient $\eta$ is more natural: we do not
need to know anything about the structure of the solution beforehand;
a fact which, for instance, allowed us to find new kinds of solutions
in~\cite{KL}. In~\cite{RMP,rm2},
the velocity is fixed by giving the ratio of the hydrodynamical and
thermal fluxes at the deflagration front.

On the other hand, the code in~\cite{RMP,rm2}
contains features we do not have. First, RMP
have an expanding background metric.
We do not expect the
extremely slow expansion to have any effect at
the short time scales we are investigating.
Second, in~\cite{rm2} the electromagnetic radiation
is treated properly, allowing RM to investigate
the decoupling of radiation from the strongly
interacting matter at a radius of about $10^4$ fm, which is important for
concentrating baryon number inside the quark droplet.
Here we study only the final stages when the bubble
has a radius smaller then 200 fm, so that the decoupling
has already taken place. Hence our number of degrees of freedom
is that of the strongly interacting particles. We
take the physical result of the decoupling period,
namely a strongly increased baryon number density within a
radius of $10^3-10^4$ fm around the droplet~\cite{rm2},
as an initial condition for our analysis of
droplet decay.

\section{Bubble growth mechanisms}
\label{bubble}

In~\cite{KL} we discussed the different hydrodynamical
growth mechanisms of phase transition bubbles.  We found three different
classes:  (1) weak deflagrations moving at a subsonic speed and preceded by a
shock front and a compression wave;  (2) Jouguet deflagrations moving at a
supersonic speed, preceded by a shock front and a compression wave, and
followed by a rarefaction wave;  and (3) weak detonations
moving at a supersonic speed, and followed by a rarefaction wave.

Note that for small supercooling, the condition of non-negative
entropy production allows only weak deflagrations~\cite{GKKM,EIKR,IKKL2}.
Moreover, adding information about the microscopic entropy producing
mechanism,  the solution becomes fixed; for instance, for the EW case
mildly relativistic weak deflagrations seem probable~\cite{mp2}.
The above three classes refer to a case when the entropy
condition does not forbid any of the solutions, and the parameter $\eta$
is allowed to vary freely. At least for the QCD case, there is ample
parameter space for all three kinds of solutions~\cite{IKKL2}.

\begin{figure}[tbh]
\vspace*{-3.0cm}
\hspace*{-1.2cm}
\epsfysize=14.0cm
\epsffile{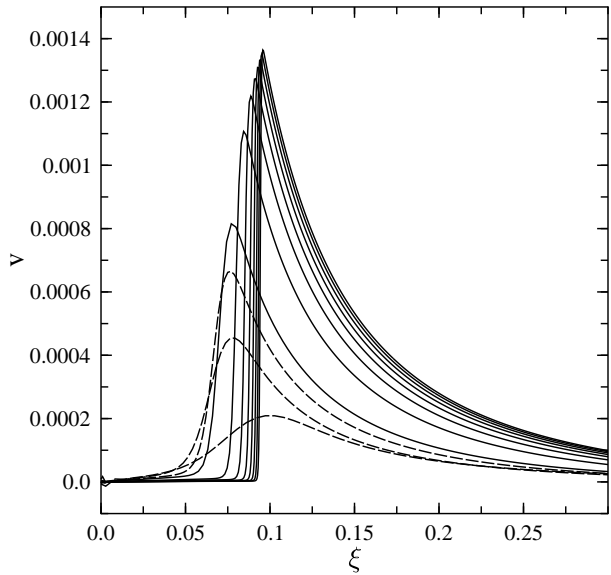}
\vspace*{-2.5cm}

\caption[a]{\protect
The velocity profile approaches the similarity solution as the bubble grows.
This run is for $\eta = 1.0T_c$.  The time interval between the dashed
curves is $\Delta t = 158.75 T_c^{-1}$, and between the solid curves
$\Delta t = 675 T_c^{-1}$.
The horizontal axis is $\xi \equiv r/t$, distance scaled with time.
}
\label{bubdev}
\bigskip
\end{figure}

We set out to verify that all three classes can be realized.
We choose parameter values which allow the three classes:  $L = 0.1T_c^4$,
$\sigma = 0.1T_c^3$, $l_c = 6T_c^{-1}$, $a = 34.25\pi^2/90$. The
nucleation temperature corresponding to these is $T_{\rm init} =
0.86T_c$~\cite{IKKL2}.
We start with initial data where there is a small ``recently
nucleated'' bubble surrounded by homogeneous fluid at rest.  The bubble will
start to grow and we follow it until the configuration becomes self-similar,
see Fig.~\ref{bubdev}.

\begin{figure}[tb]
\vspace*{-3.0cm}
\hspace*{-1.2cm}
\epsfysize=14.0cm
\epsffile{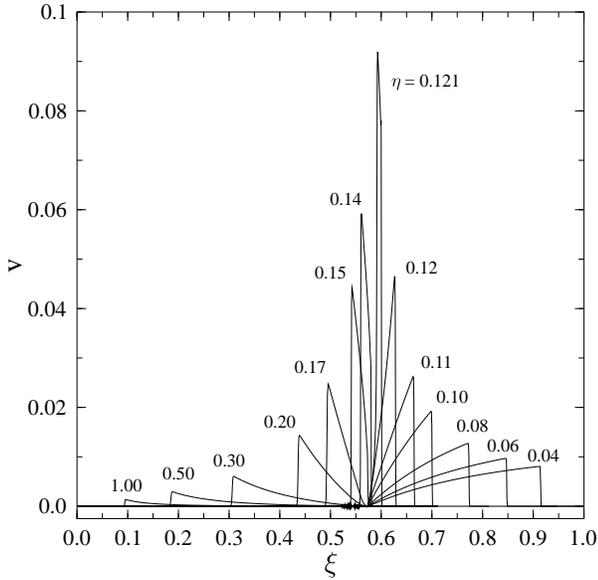}
\vspace*{-2.5cm}

\caption[a]{\protect
A sequence of velocity profiles for bubbles with different values of $\eta$
(given in units of $T_c$ next to each profile).
These profiles are from our hydrodynamical runs.
Compare with Fig.~3 of~\cite{KL}, where the profiles are solutions of
eqs.~(\ref{veq}) and~(\ref{Teq}) (with a bag equation of state).
}
\label{xiset}
\bigskip
\end{figure}

\begin{figure}[tbh]
\vspace*{-3.0cm}
\hspace*{-1.2cm}
\epsfysize=14.0cm
\epsffile{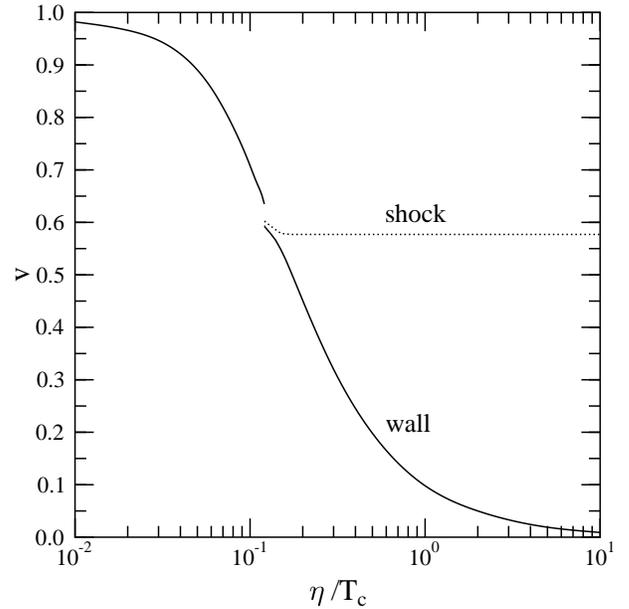}
\vspace*{-2.5cm}

\caption[a]{\protect
The bubble wall velocity and the shock velocity as a function of $\eta$.
}
\label{bubeta}
\bigskip
\end{figure}

\begin{figure}[p]
\vspace*{-3.0cm}
\hspace*{-1.2cm}
\epsfysize=12.0cm
\epsffile{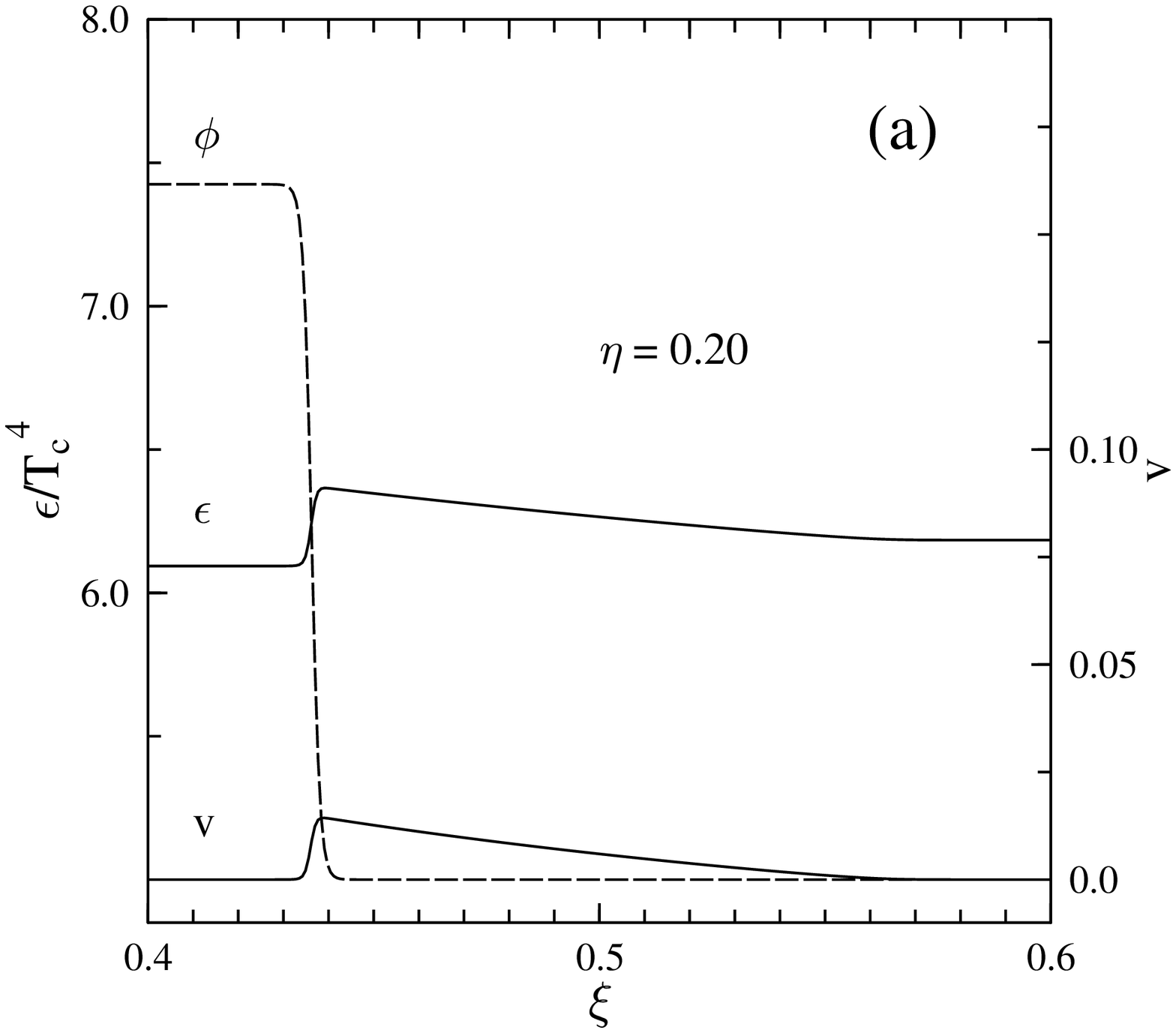}
\vspace*{-4.0cm}

\vspace*{-1.0cm}
\hspace*{-1.2cm}
\epsfysize=12.0cm
\epsffile{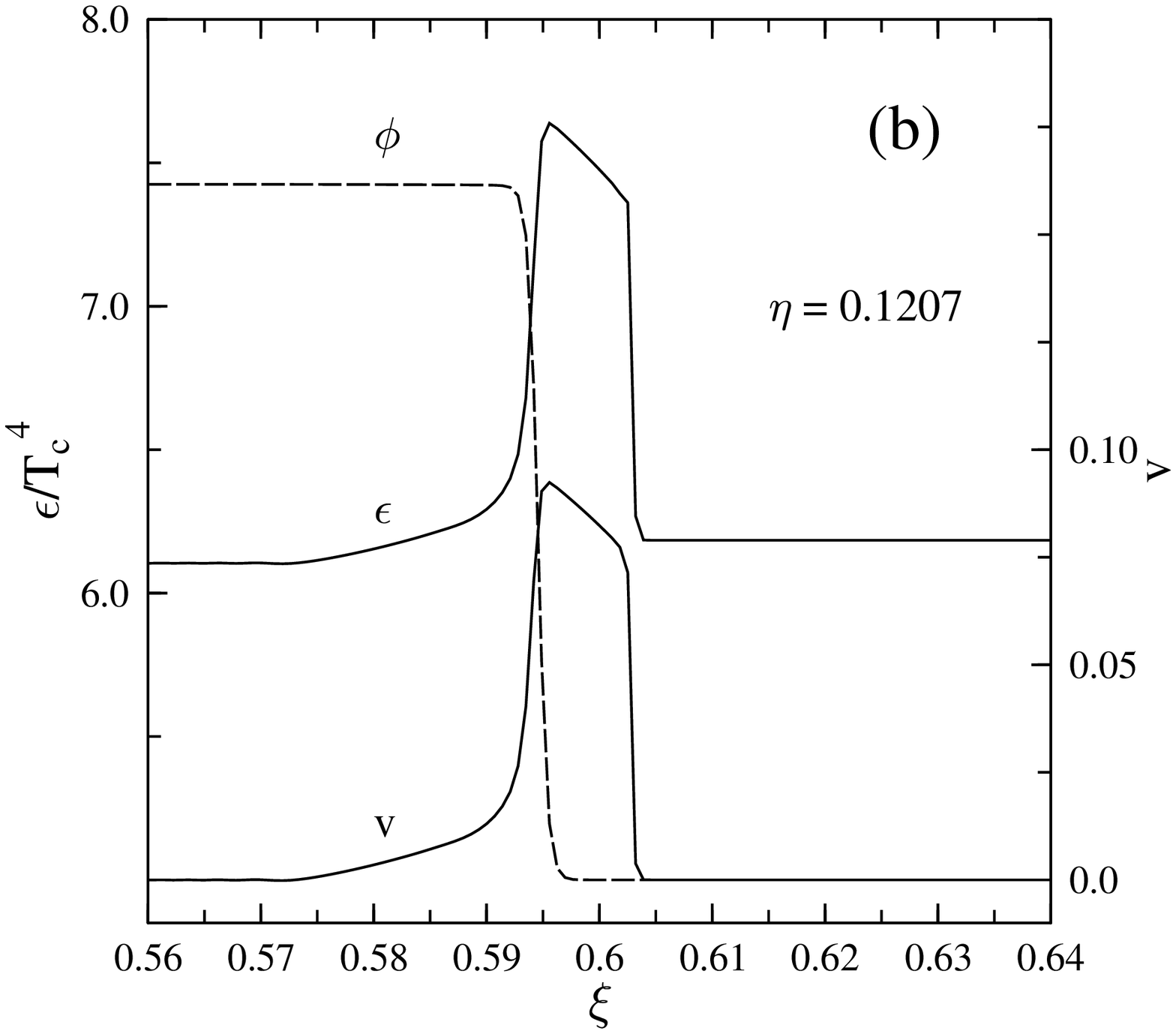}
\vspace*{-4.0cm}

\vspace*{-1.0cm}
\hspace*{-1.2cm}
\epsfysize=12.0cm
\epsffile{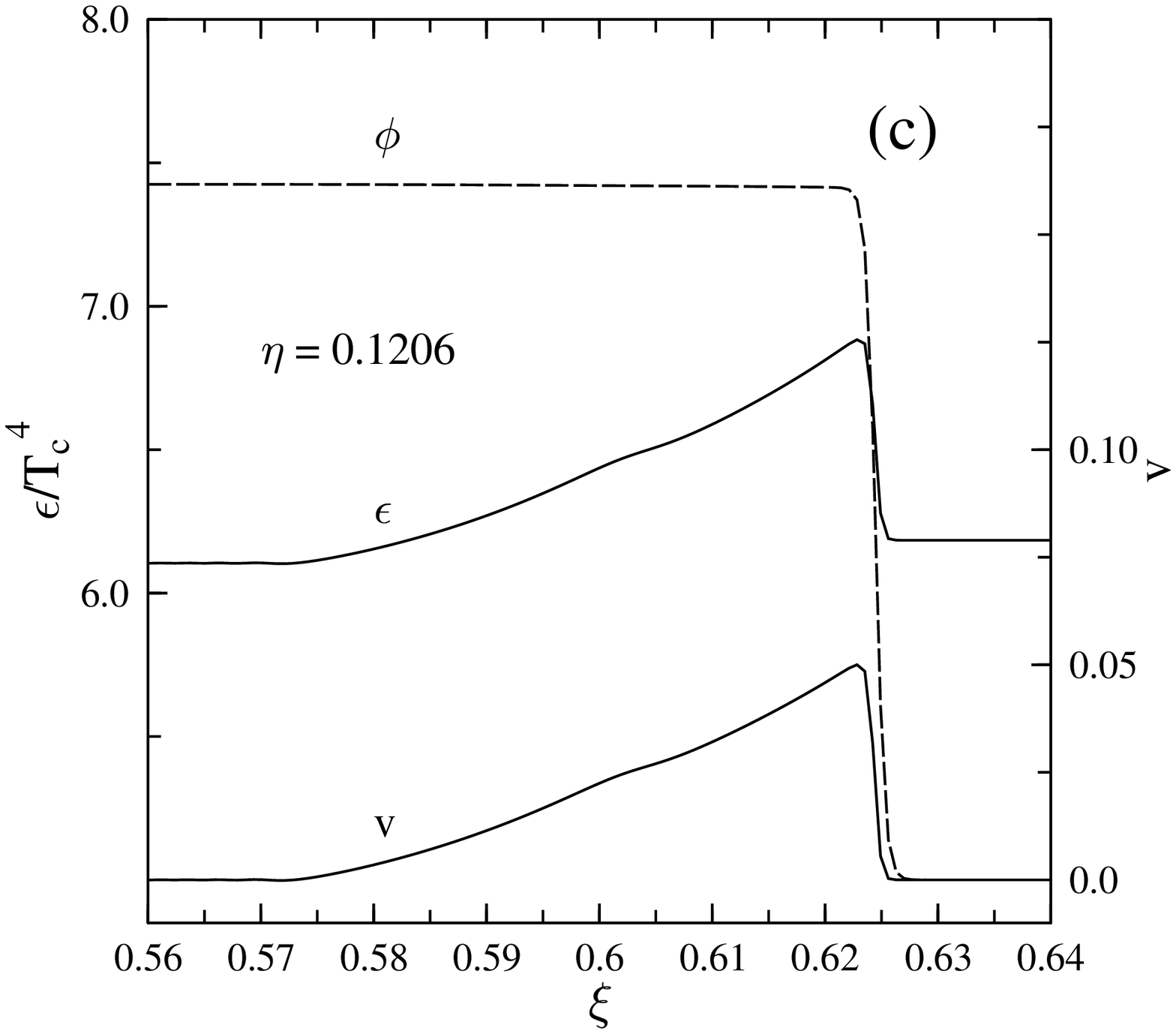}
\vspace*{-2.5cm}

\caption[a]{\protect
The energy density, velocity, and order parameter profiles for the three
different solution classes. The first (a) is a weak deflagration,
the second (b) a Jouguet deflagration,
and the third (c) a weak detonation.
}
\label{suplot}
\bigskip
\end{figure}

The velocity of the bubble wall will depend on the dissipative constant $\eta$.
We covered the range from $\eta = 0.01T_c$ to $\eta = 10T_c$, and found all
three classes:  (1) weak deflagrations for $\eta \gtrsim 0.13T_c$;  (2) Jouguet
deflagrations for $0.12T_c \lesssim \eta \lesssim 0.13T_c$;
and (3) weak detonations
for $\eta \lesssim 0.12T_c$.  In Fig.~\ref{xiset} we show a sequence of final
velocity profiles for a set of runs, and in Fig.~\ref{bubeta}
we show the wall and
shock velocities as a function of $\eta$.  The shock fronting the compression
wave in the deflagration bubble solutions becomes exponentially weak for
slowly growing bubbles,
and we did not resolve the shock for $\eta \gtrsim 0.17T_c$.
Fig.~\ref{suplot} shows an example from each solution class.

We see (Fig.~\ref{bubeta})
that as $\eta$ is decreased the wall velocity grows smoothly until it is
close to the speed of sound.  Then the wall velocity appears to have some
difficulty breaking the sound barrier.  (In a Jouguet  deflagration the wall
is subsonic with respect to the fluid just ahead of the wall, and it is
moving at the speed of sound with respect to the fluid just behind, although
the velocity with respect to the
bubble center, ``the wall velocity'', already
exceeds the speed of sound).
The increase in wall velocity slows
down, until at $\eta \sim 0.12T_c$ the solution shifts from a deflagration to a
detonation solution and the wall velocity jumps abruptly.  Thus we do not have
solutions for wall velocities between $v = 0.593$ and $v = 0.635$.

\begin{figure}[tbh]
\vspace*{-3.0cm}
\hspace*{-1.2cm}
\epsfysize=14.0cm
\epsffile{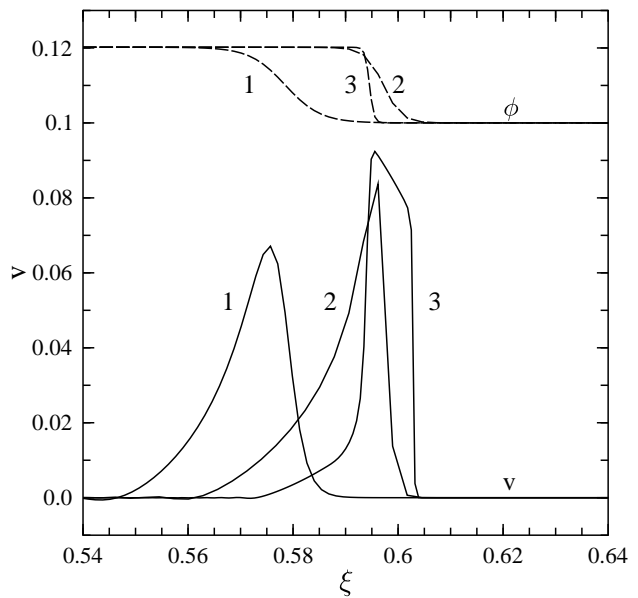}
\vspace*{-2.5cm}

\caption[a]{\protect
A bubble ($\eta = 0.1207T_c$) which first goes to a detonation configuration
before settling to a deflagration.  The three profiles are from times $t_1 =
1350T_c^{-1}$, $t_2 =  2700T_c^{-1}$, and $t_3 = 10800T_c^{-1}$.
}
\label{detotodefl}
\bigskip
\end{figure}

In~\cite{KL} we described similarity solutions for all wall velocities.  The
reason that we do not here find solutions in the above range
$0.593\ldots 0.635$, is that there are
pairs of solutions with different wall velocities corresponding to the same
value of $\eta$.  The slower one is a Jouguet deflagration, the faster one a
weak detonation.  Only one of these two is realized for a given
initial configuration.  Thus when the solution
shifts from a deflagration to a detonation, there is a jump in wall velocity.
In fact, in the runs with the fastest deflagrations, the bubble first went into
a detonation configuration before settling into a deflagration, see
Fig.~\ref{detotodefl}.

\section{Droplet decay}
\label{droplet}

Kajantie and Kurki-Suonio (KK)~\cite{KajKur} have discussed the shrinking quark
droplet.  There is an outward fluid flow from the surface of the droplet.  When
the droplet vanishes the source of this flow disappears.  This led KK to
conclude that the evaporation site sends out a rarefaction wave by which the
outward flow is stopped.  KK did not consider the effect of surface tension on
droplet evaporation.

Rezzolla, Miller, and Pantano~\cite{RMP} have improved upon the work of
KK.  They conclude that the flow pattern around a shrinking droplet settles
into a similarity solution.  If the similarity solution would hold till the
end, the droplet would leave behind a homogeneous fluid at rest, since all
structure around the droplet would have shrunk to zero size at the moment the
droplet vanishes.  But as the droplet becomes small, the surface tension
causes the droplet wall to accelerate, and the droplet actually vanishes before
the surrounding outward flow pattern.  Apparently RMP have not followed the
evolution past the droplet disappearance with their code.  This we have done
with ours.

We start with the similarity solution as an initial condition.  Thus we need to
construct this solution in our model. Let us first discuss the regions
away from the phase transition surface,
where the order parameter lies at the minima
$\phi = 0$ and $\phi = \phi_{\rm min}(T)$ in the two phases
($\partial V/\partial\phi = 0$). We
assume the distance scale is large enough
so that $\partial_\mu\phi$ terms can be
ignored in $T^{\mu\nu}$.  Then we have
\begin{eqnarray}
\epsilon_q(T) & = & 3aT^4, \\
p_q(T)        & = & aT^4, \\
\epsilon_h(T) & = & 3aT^4 + V(\phi_{\rm min}(T),T) -
T\frac{\partial V}{\partial T}
\bigl(\phi_{\rm min}(T),T\bigr), \\
p_h(T)        & = & aT^4 - V\bigl(\phi_{\rm min}(T),T\bigr).
\end{eqnarray}

\begin{figure}[tbh]
\vspace*{-3.0cm}
\hspace*{-1.2cm}
\epsfysize=14.0cm
\epsffile{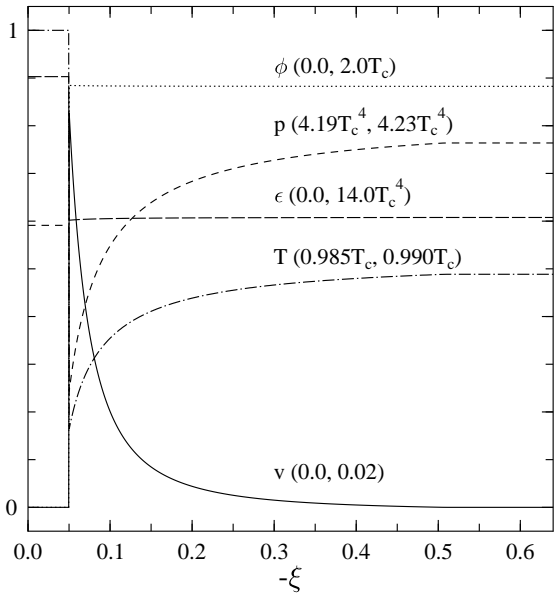}
\vspace*{-2.5cm}

\caption[a]{\protect
A similarity solution for a shrinking droplet.
The two values in the brackets refer to the value of the
corresponding quantity at ``0'' and ``1'' of the $y$-axis.
}
\label{simsol}
\bigskip
\end{figure}

The similarity solution depends on the coordinates $r$ and $t$ only through the
combination $\xi \equiv r/t$.  For a shrinking solution, we must choose $t = 0$
at the moment of disappearance,  so $t < 0$ and thus $\xi < 0$.  The similarity
solution has homogeneous fluid at rest inside the droplet, $v = 0$, $\epsilon =
\epsilon_q(T_q)$.  The droplet
surface at $\xi = \xi_{\rm defl}$ is a deflagration front,
the phase boundary moving
in at a speed $|\xi_{\rm defl}|$.  The usual deflagration
conditions give us then the
fluid state just outside the droplet wall, with fluid flowing outwards.  The
profile $\epsilon(\xi)$, $v(\xi)$ of this compression wave
is solved from~\cite{St,KS}
\begin{eqnarray}
\biggl[\frac{1}{c_s^2(T)}(v-\xi)^2 - (1-v\xi)^2\biggr]\frac{dv}{d\xi} & = &
   2\frac{v}{\xi}(1-v\xi)(1-v^2), \label{veq}\\
\frac{1}{\epsilon(T)+p(T)} \frac{dp}{dT}(T)\frac{dT}{d\xi} & = &
   \frac{\xi-v}{1-v\xi}\frac{1}{1-v^2} \frac{dv}{d\xi}.
\label{Teq}
\end{eqnarray}
The velocity is positive (outwards) and becomes zero at the sonic point
$\xi = -c_s$, where we have a weak discontinuity.  For $|\xi|>c_s$ the fluid is
at rest with constant energy density.  (Note that there is no shock front,
unlike in the case of an expanding deflagration bubble).  In Fig.~\ref{simsol}
we show a similarity solution for
the case $T_q = 0.99$, $\xi_{\rm defl} = -0.05$.

For the phase transition surface we use a finite-width wall
obtained with our stationary plane-symmetric
code~\cite{IKKL}. This gives the initial data for our
dynamical code.

We present results from a run with parameter
values $L = 4T_c^4$, $\sigma = 0.5
T_c^3$, $l_c = 1T_c^{-1}$, and $a = 40\pi^2/90$, using an initial condition
with $T_q = 0.99T_c$.  (This is a fairly arbitrary choice, as there is large
uncertainty about the values of these parameters, and we are thus looking for
qualitative features only).
The initial data is the similarity solution for the case
$v_{\rm defl} = -0.05$, and we set the initial
radius of the droplet to be $r = 200
T_c^{-1}$ (for QCD $T_c^{-1}$ is just 1 fm,
and for the EW case $10^{-3}$ fm).
Now the actual wall velocity will depend on the dissipation
parameter $\eta$.  The value of $\eta$ corresponding to a given $v_{\rm defl}$
can be found by trying out different values.  Because of the fairly small
initial size, there are already some small-scale effects and the solution
should be deviating somewhat from the similarity solution.  Therefore
using the similarity solution as initial data will cause some
disturbance to spread out from the wall to alter the profile.
We minimized this
initial disturbance by choosing $\eta = 1.97 T_c$, giving an actual initial
wall velocity $v_{\rm defl} = -0.047$.

\begin{figure}[p]
\vspace*{-0.4cm}
\hspace*{-1.2cm}
\epsfysize=14.0cm
\epsffile{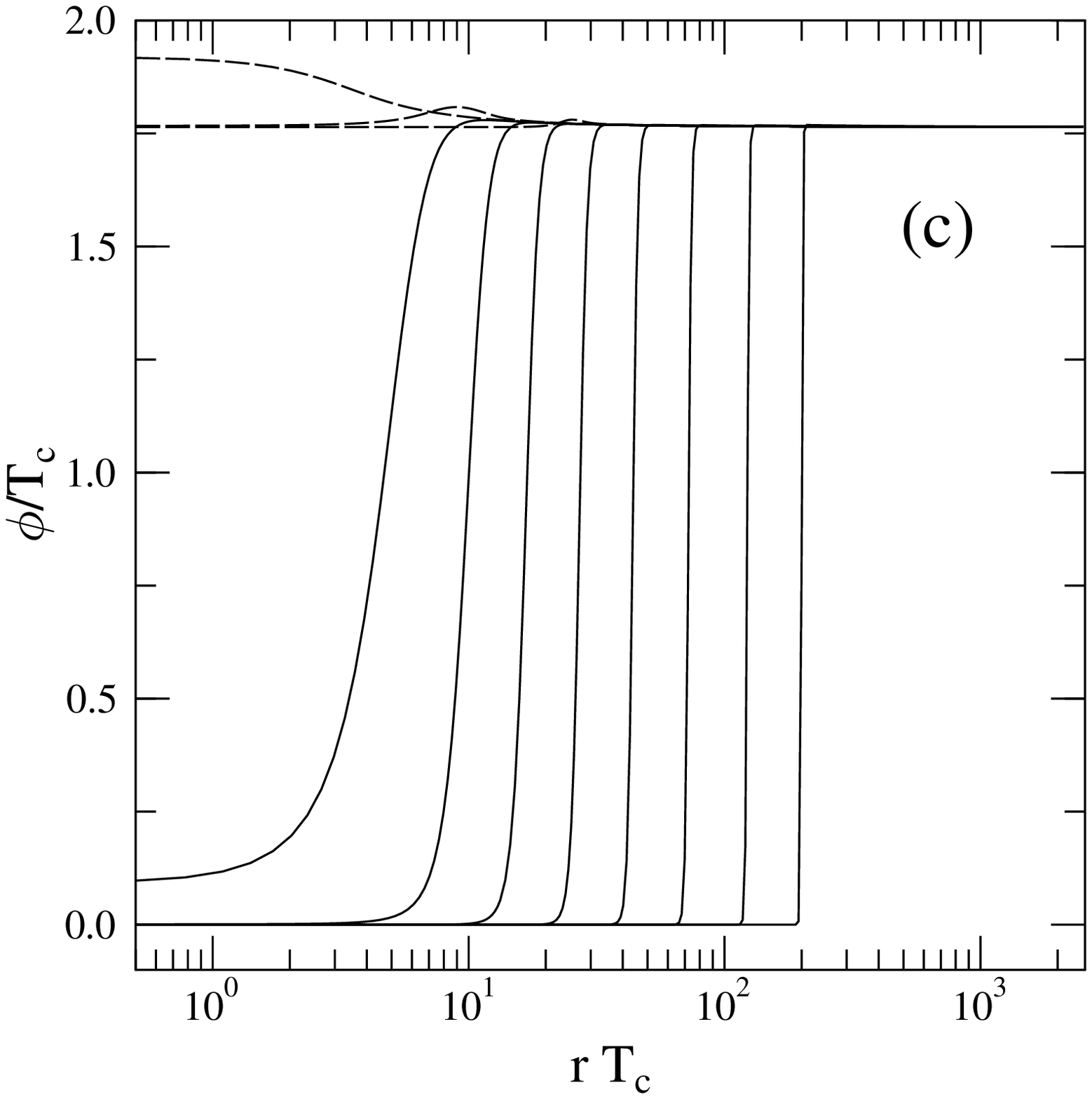}

\vspace*{-4.0cm}

\vspace*{-1.0cm}
\hspace*{-1.2cm}
\epsfysize=14.0cm
\epsffile{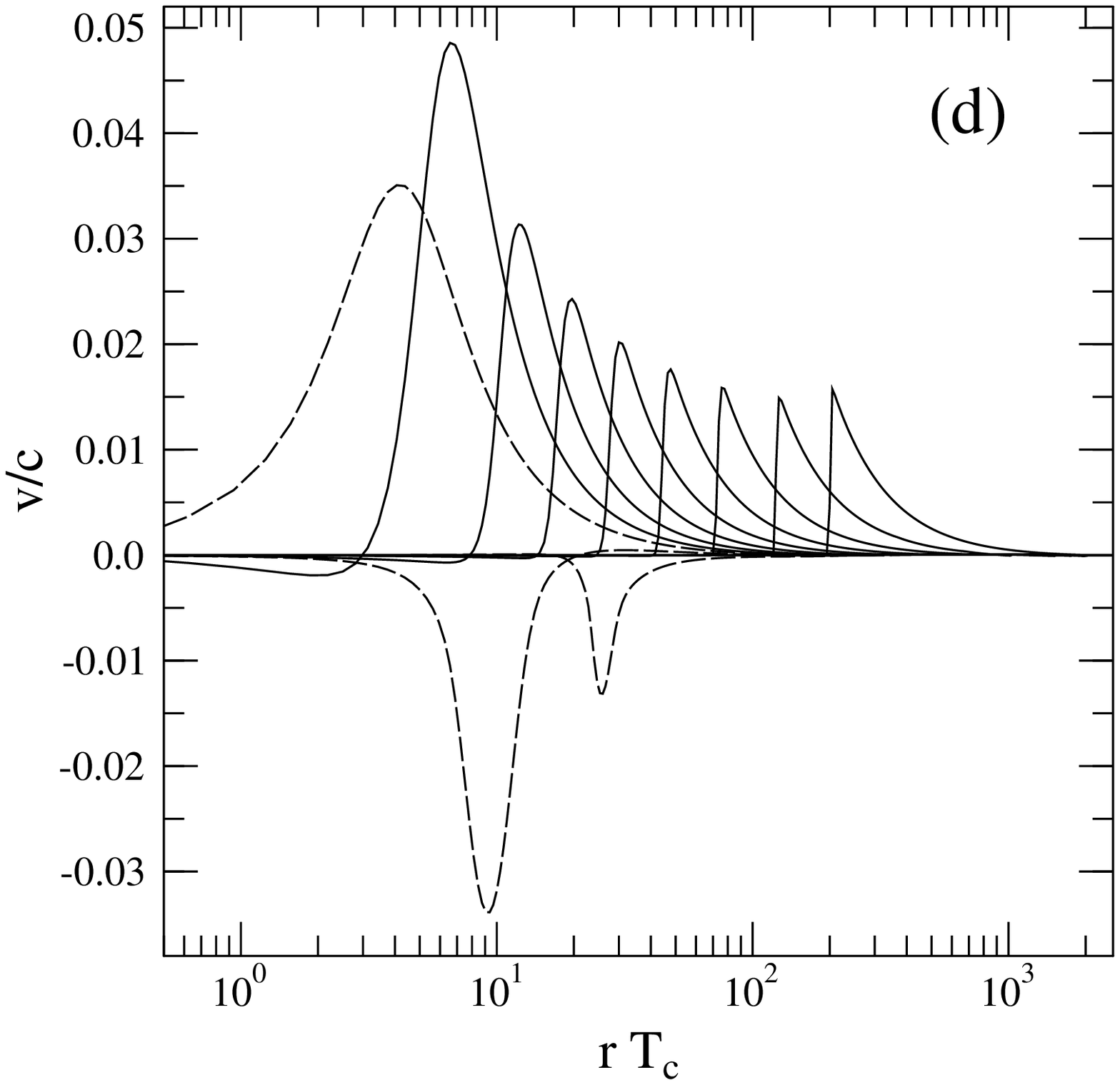}

\vspace*{-2.5cm}

\caption[a]{\protect
Results from the run.  We show profiles for four quantities:
proper energy density (a), local temperature (b),
order parameter (c), and
flow velocity (d).  The solid curves are from successive time-slices as the
droplet is shrinking (wall moving to the left).  The dashed curves
are from time-slices after the droplet has disappeared (pattern moving to the
right).  We use a logarithmic axis for the radial coordinate to focus on the
droplet evaporation site.
The finite thickness of the wall becomes thus apparent
when the wall has reached small values of $r$.
}
\label{run}
\bigskip
\end{figure}

\begin{figure}[p]
\vspace*{-1.0cm}
\hspace*{-1.2cm}
\epsfysize=14.0cm
\epsffile{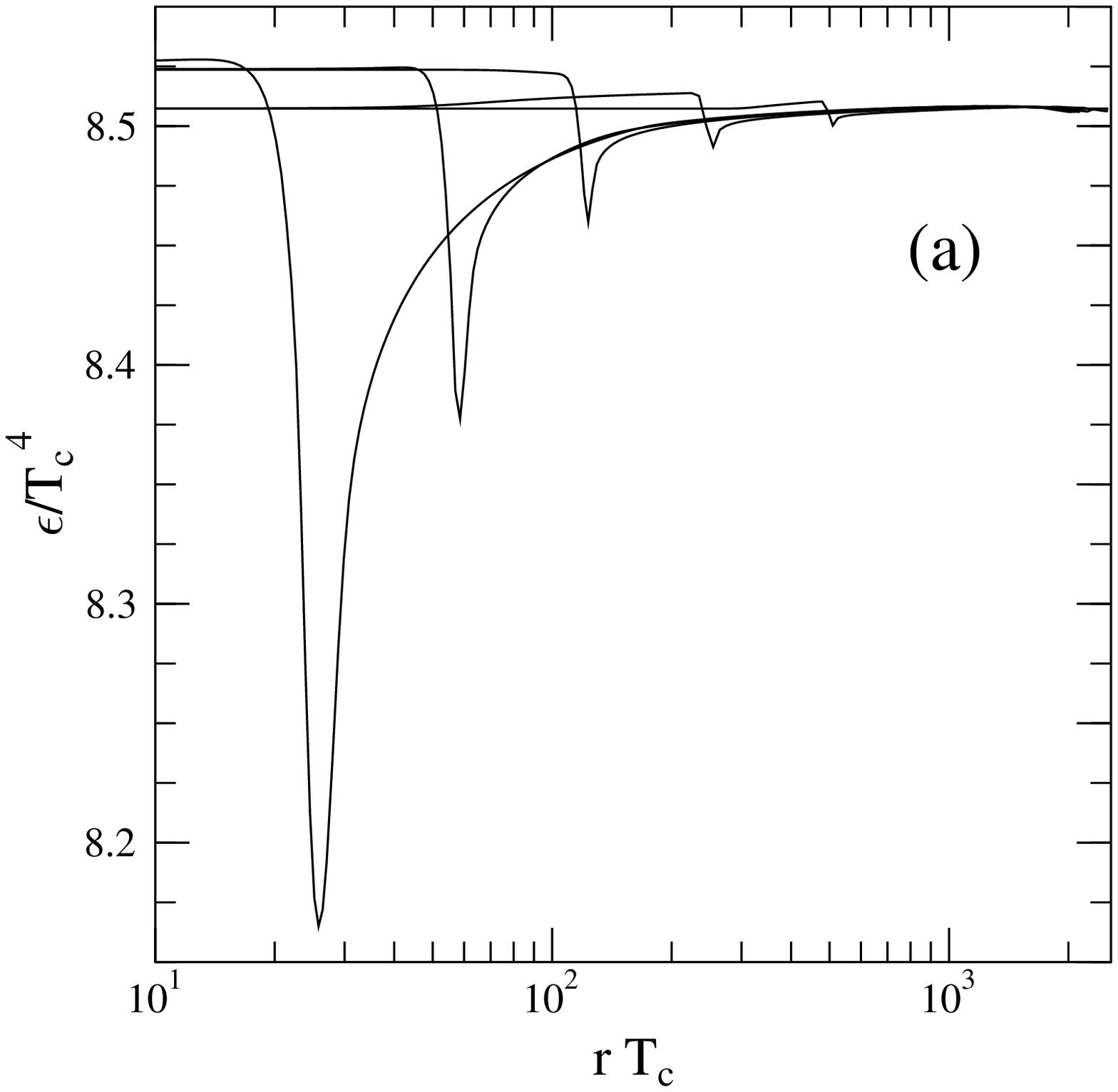}

\vspace*{-5.0cm}
\hspace*{-1.2cm}
\epsfysize=14.0cm
\epsffile{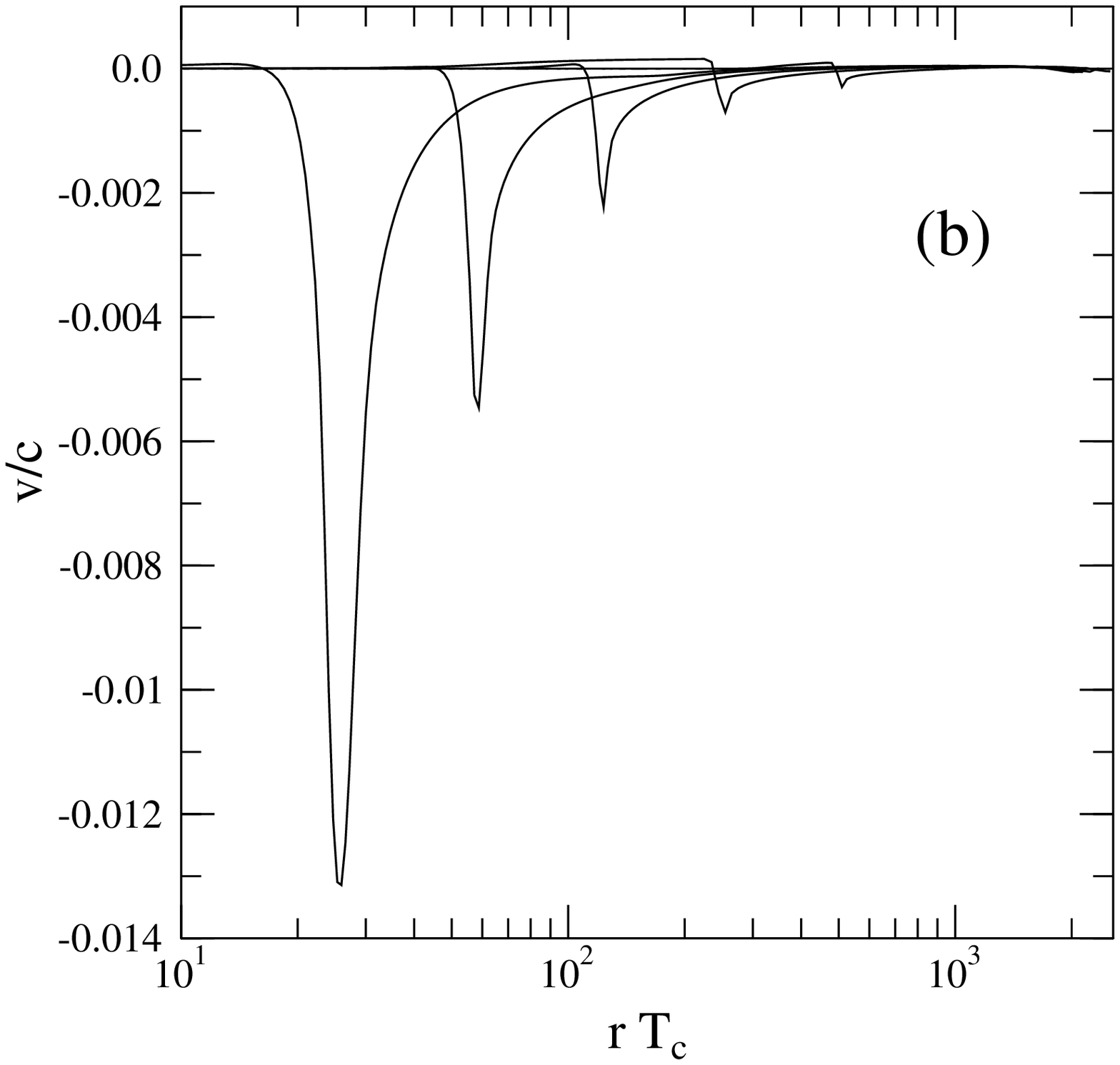}

\vspace*{-2.5cm}

\caption[a]{\protect
Results from the same run as Fig.~\ref{run}.
We show later time slices (after droplet evaporation) of proper energy density
(a) and flow velocity (b) with an expanded scale.
A pulse of inward flow is moving outwards and decaying rapidly, evening out the
energy density (and other thermodynamic quantities).
}
\label{runb}
\bigskip
\end{figure}

\begin{figure}[p]
\vspace*{-2.4cm}
\hspace*{-1.2cm}
\epsfysize=14.0cm
\epsffile{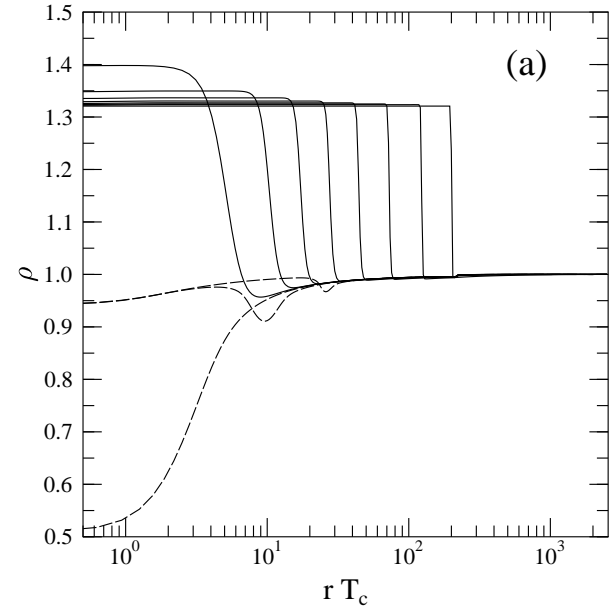}

\vspace*{-5.0cm}
\hspace*{-1.2cm}
\epsfysize=14.0cm
\epsffile{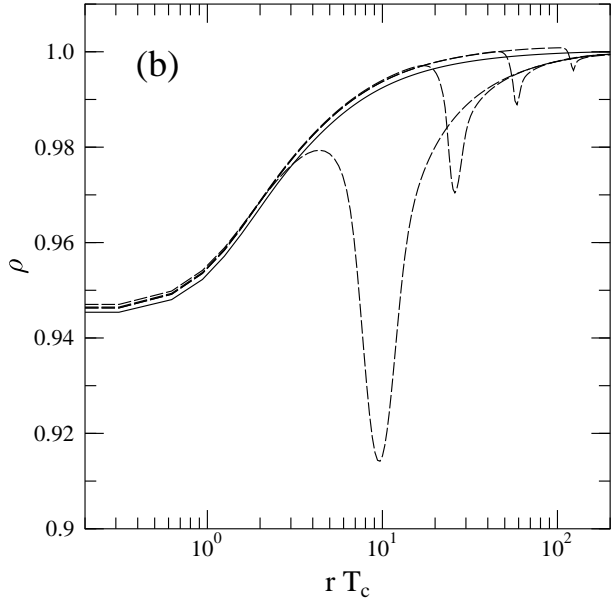}

\vspace*{-2.5cm}
\caption[a]{\protect
The compression factor $\rho$.
The upper figure (a) is like Fig.~\ref{run}.
The lower figure (b) is a close-up
near the droplet evaporation site and shows time slices after droplet
evaporation.  The solid line is the final $\rho$-profile, showing the
decompression due to droplet surface energy.
Note the very small spatial extension of the decompressed region.
}
\label{rho}
\bigskip
\end{figure}

\begin{figure}[p]
\vspace*{0.0cm}
\hspace*{-1.2cm}
\epsfysize=14.0cm
\epsffile{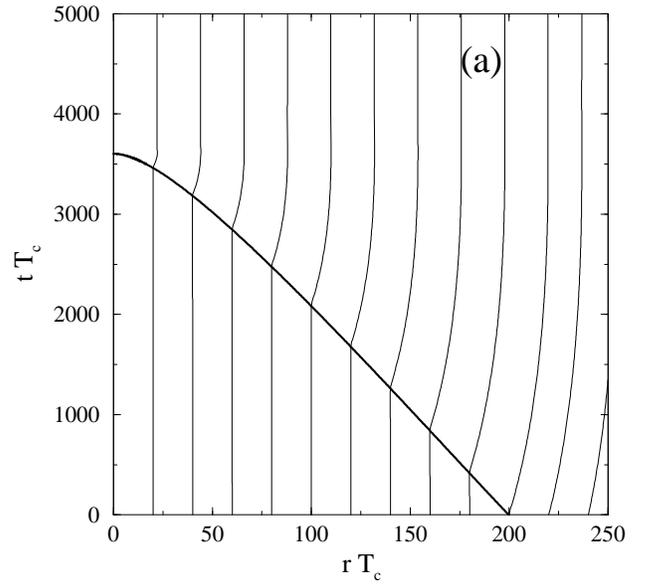}

\vspace*{-5.0cm}
\hspace*{-1.2cm}
\epsfysize=14.0cm
\epsffile{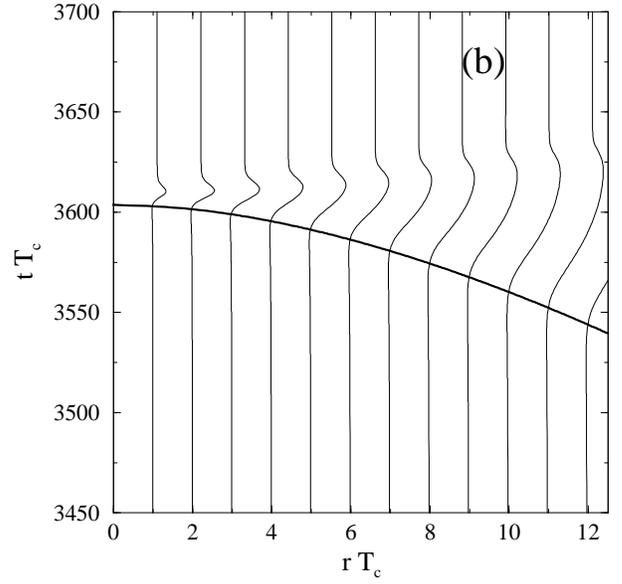}

\vspace*{-4.5cm}

\caption[a]{\protect
The flow lines of particles moving with the fluid.
The lower figure (b) is a magnification of the final stages.
This figure contains the same information as Fig.~\ref{rho},
but in a different form.
At the time of the droplet decay, there is a local rarefaction and
subsequent compression near the droplet, but no global
rarefaction wave is left behind.
Due to the difficulty of getting baryon number across the phase transition
surface (the thick line),
a part of the baryon number would not follow the flow lines
but would instead remain in the quark phase.
}
\label{flow}
\bigskip
\end{figure}

Figs.~\ref{run}--\ref{flow} show results from this run.  At first we stay
fairly close to the similarity solution but as the droplet gets
smaller the wall is accelerated inwards, and the temperature and energy density
rise inside the droplet.  As the droplet disappears the outward flow leads to
an underdense region with lower temperature in the center.  The fluid flow then
turns around filling this underdense region.  A pulse of inward flow is thus
sent outwards, and decays rapidly.  The end result is homogeneous fluid at
rest.
Thus we see no rarefaction wave
propagating to infinity after the decay. This
is in contrast to expectations in~\cite{KajKur,RMP,rm2}.

In Fig.~\ref{rho} we show the compression factor $\rho$ (density of
noninteracting test particles moving with the fluid).  In the end the
thermodynamical properties ($T,v,\phi$) of the fluid become
homogeneous.
There is a small depression in the final
$\rho$ at the droplet evaporation site.  This is due to the droplet surface
energy which was converted into
thermal energy.  Because of this extra energy some
decompression was needed to even out the energy density.

Note that
due to the difficulty of getting baryon number through the wall, the baryon
number tends to concentrate on the inside of the droplet surface, and is thus
not moving with the fluid~\cite{w,AppHog,FMA,KS2,SKAM,JF2,rm2}.
Thus the compression factor $\rho$ does not represent baryon
number density.  However, it can be viewed as representing a background motion
of the baryon number, upon which the baryon-number-concentrating effects are
superimposed.  Thus we see that the hydrodynamical flow itself does not lead
to any inhomogeneity in the  baryon number, except the tiny depression at the
droplet evaporation site.

RM~\cite{rm2} studied the effect of radiative energy transfer out of the quark
droplet on the baryon number distribution.   We are studying scales where this
transfer has ceased to operate as the droplet has become transparent to
radiation.  Thus we can combine our work with RM by considering their result
at the relevant time as an initial condition for our runs.  Their compression
factor $\rho_{\rm RM}$ can be viewed as being proportional to baryon number
density in an idealization where other baryon number concentrating effects are
ignored.  Let us do this idealization in the discussion below.  RM found that
the shrinking droplet leaves behind a region with a high overdensity in baryon
number, whose radius is of the order of the mean free path $\lambda \sim 10^4
{\rm fm}$ of the electromagnetic radiation.  Inside the quark droplet there is
an even higher baryon number density.  As we are
considering the region $r \ll \lambda$, we can take our compression factor
$\rho$ to denote the baryon number density normalized to the value it has in
the end at about $r \sim 10^2 {\rm fm}$, where it is fairly homogenous.
Thus our Fig.~\ref{rho}(a) shows the baryon number density contrast between
the quark and hadron phases (the numerical value is much smaller in our case
than in~\cite{rm2} because of different parameter values used).
There are two contributions to
this density contrast: (1) the decompression due to the change in equation of
state as the fluid passes through the phase boundary, and (2) the extra
overdensity inside the droplet due to surface tension, which becomes
significant in the end.  There will be no remnant of this higher density
inside the droplet after the droplet has disappeared (in reality, there
probably will be a remnant due to effects we are ignoring here~\cite{KS2}).
Contribution (1) is eliminated as the fluid is converted to the hadron phase.
Contribution (2) is eliminated as the thermodynamical variables are evened out
by the outward pulse (see Fig.~\ref{runb}).  Thus the $\rho_{\rm RM}$ outside
the droplet represents the final baryon number distribution, and it is not
disturbed by any rarefaction wave.  For $r \ll \lambda$ the baryon number
density is homogeneous except for the central small depression caused by the
released droplet surface energy (see Fig.~\ref{rho}(b)).  The baryon number
overdensity within $r \lesssim 10^4 {\rm fm}$ is later eliminated mainly by
neutron diffusion~\cite{JF1}.

\section{Conclusions}
\label{concl}

We have studied the growth of bubbles and decay of droplets in cosmological
phase transitions using a spherically symmetric version of our hydrodynamic
code based on a cosmic-fluid--order-parameter model~\cite{IKKL} of the
transition.

We demonstrate how an initial small newly nucleated bubble begins to grow and
evolves to a similarity solution.  The wall velocity depends on the dissipative
constant $\eta$.  We find all three classes of hydrodynamic
similarity solutions: weak deflagrations, Jouguet deflagrations, and weak
detonations.  In the first class the wall velocity is subsonic, in the other
two classes supersonic with respect to the origin.

Not all wall velocities are realized as the value of $\eta$ is varied.  For a
certain range in $\eta$, there would be both a Jouguet deflagration and a weak
detonation solution with the same value of $\eta$, but a different wall
velocity.  As the solution shifts from the former class to the latter, there
is thus a jump in wall velocity.

Note that all classes of solutions are not possible for all values of the other
parameters~\cite{GKKM,EIKR,IKKL2}.
Typically weak deflagrations are always possible,
but detonations and Jouguet deflagrations may not be
reached; or the solution may shift directly from weak deflagrations to weak
detonations as $\eta$ is decreased, skipping the class of Jouguet
deflagrations~\cite{IKKL2}.

We followed the evaporation of a quark droplet starting from a similarity
solution.  The droplet deviates from the similarity solution in the end when
small-scale effects from the surface tension and finite wall thickness become
significant.  The nature of the fluid flow after the droplet evaporation
differs from what has been previously suggested~\cite{KajKur,RMP,rm2}. The
deviation from similarity flow in the end leads to a temporary energy
overdensity at
the evaporation site, surrounded by a shell of underdensity.  This
inhomogeneity is eliminated by an outward moving pulse which decays rapidly.
This effect does not extend beyond the small scales where the droplet
deviated from the similarity solution.

Our model does not contain the effects which concentrate baryon number:
the suppression of baryon flow through the wall and the radiative entropy
transfer out of the droplet~\cite{w,AppHog,FMA,KS2,SKAM,rm2}.
The temporary hydrodynamic
compression at the final stage of droplet decay does not
affect the baryon/entropy ratio
and thus has no lasting effect on the baryon density.  There is a small
decompression at the very center to accommodate the droplet surface energy
which has been converted to thermal energy. The
absence of a global rarefaction wave indicates that any previously
generated baryon number inhomogeneity
at the scale $10^4$ fm is not diluted away
by the hydrodynamics related to droplet decay.

\section*{Acknowledgements}

We are grateful to the
Center for Scientific Computing (Finland) for computational resources.

\begin{figure}[p]
\vspace*{-3.0cm}
\hspace*{-1.2cm}
\epsfysize=14.0cm
\epsffile{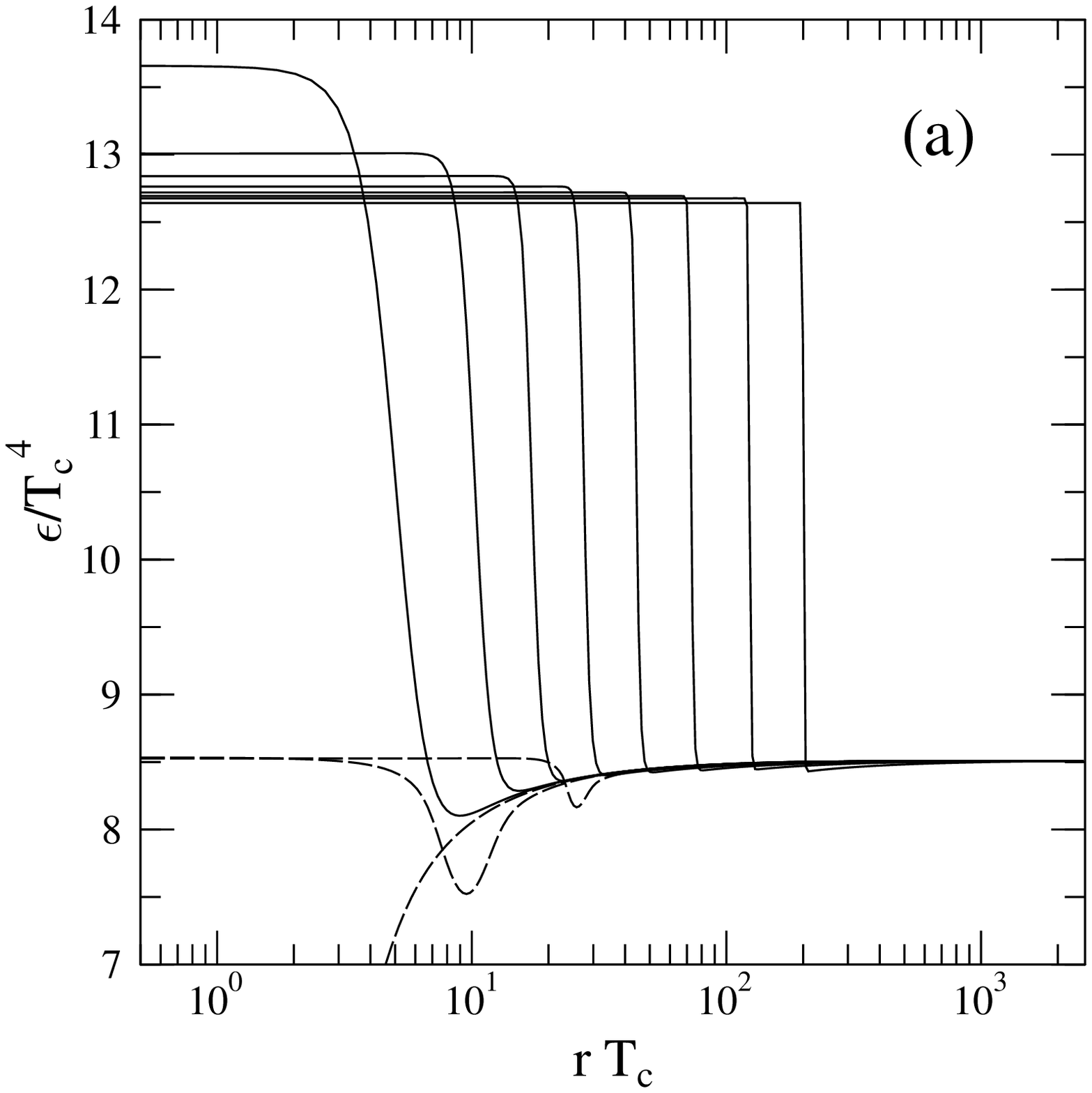}

\vspace*{-4.0cm}

\vspace*{-1.0cm}
\hspace*{-1.2cm}
\epsfysize=14.0cm
\epsffile{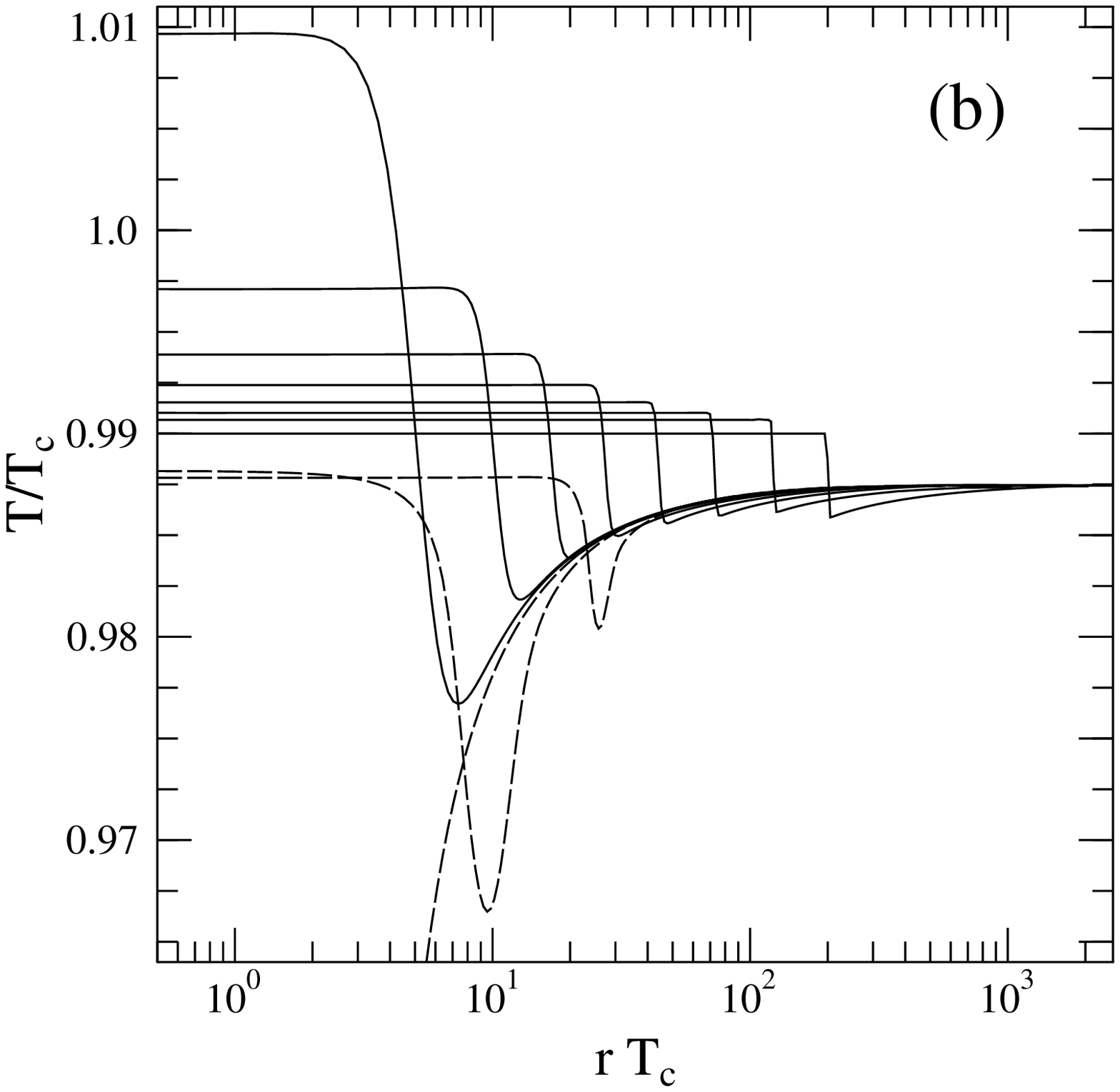}

\vspace*{-1.0cm}
\end{figure}

\end{document}